\documentclass[twocolumn,iop,numberedappendix,appendixfloats]{oja}

\usepackage[utf8]{inputenc}
\usepackage[british]{babel}

\usepackage{graphicx}	
\usepackage{amsmath}	
 \usepackage{amssymb}	
 \usepackage{nicefrac}
\usepackage{subcaption}  
\usepackage{caption}
\usepackage{xcolor}
\definecolor{maroon}{cmyk}{0, 0.87, 0.68, 0.32}
\definecolor{halfgray}{gray}{0.55}
\usepackage{float}
\usepackage{fontawesome}
\usepackage[colorlinks,allcolors=blue]{hyperref}
\usepackage{dcolumn}
\usepackage{bm}
\usepackage{physics}
\usepackage{xspace}
\usepackage{newtxtext,newtxmath}

\usepackage{url}

\newlength{\twocolwidth}

\usepackage{array,multirow}
\usepackage{ulem}

\providecommand{\dd}{\ensuremath{{\rm d}}}

\newcommand{\mPsi}{{\mathit \Psi}}
\renewcommand{\vec}[1]{\ensuremath{\boldsymbol{#1}}}

\definecolor{mygreen}{RGB}{20,138,6}
\definecolor{myblue}{RGB}{52,180,230}
\definecolor{mypurple}{RGB}{160,60,160}

\newcommand{\orcid}[1]{\href{https://orcid.org/#1}{\includegraphics[width=10pt]{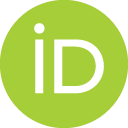}}}

\begin{document}

\journalinfo{The Open Journal of Astrophysics}
\submitted{}

\title{Stochastic super-resolution of cosmological simulations with denoising diffusion models}

\shorttitle{Diffusion models for cosmological simulations} 
\shortauthors{A. Schanz, F. List, O. Hahn}

\author{A. Schanz$^{\star1}$\orcid{0009-0001-8470-179X}} 
\author{F. List$^{1,2}$\orcid{0000-0002-3741-179X}}
\author{O. Hahn$^{1,2}$\orcid{0000-0001-9440-1152}}

\affiliation{$^1$Department of Astrophysics, University of Vienna, T\"{u}rkenschanzstra{\ss}e 17, 1180 Vienna, Austria}
\affiliation{$^2$Department of Mathematics, University of Vienna, Oskar-Morgenstern-Platz 1, 1090 Vienna, Austria}

\thanks{$^\star$ E-mail: \nolinkurl{andreas.schanz@univie.ac.at}}

\date{\today}

\begin{abstract}
In recent years, deep learning models have been successfully employed for augmenting low-resolution cosmological simulations with small-scale information, a task known as `super-resolution'. So far, these cosmological super-resolution models have relied on generative adversarial networks (GANs), which can achieve highly realistic results, but suffer from various shortcomings (e.g. low sample diversity). We introduce denoising diffusion models as a powerful generative model for super-resolving cosmic large-scale structure predictions (as a first proof-of-concept in two dimensions). To obtain accurate results down to small scales, we develop a new `filter-boosted' training approach that redistributes the importance of different scales in the pixel-wise training objective. We demonstrate that our model not only produces convincing super-resolution images and power spectra consistent at the percent level, but is also able to reproduce the diversity of small-scale features consistent with a given low-resolution simulation. This enables uncertainty quantification for the generated small-scale features, which is critical for the usefulness of such super-resolution models as a viable surrogate model for cosmic structure formation. 
\vspace*{0.05cm}
\end{abstract}

\maketitle



\section{Introduction}

A big part of the success of the $\Lambda$CDM ($\Lambda$ cold dark matter) standard model of cosmology can be ascribed to the excellent agreement between astronomical observations and the predictions made by cosmological simulations of the Universe, which follow the dynamics of matter through cosmic history. Since the first $N$-body simulations in the 1960s, which used $N \leq 100$ particles \citep{vonHoerner1960, Aarseth1963}, computational progress has been rapid, keeping pace with Moore's law: the Millennium simulation crossed the barrier of $N = 10^{10}$ particles in 2005 \citep{millennium} and more recently, simulations with more than $10^{12}$ particles have been performed (e.g.\ \citealt{potter2017pkdgrav3, OuterRim2}). While this progress is undeniably impressive, the range of scales in the Universe is so enormous that even in the recent \textsc{Uchuu} simulation \citep{Uchuu} with $N = 2.3 \times 10^{12}$ particles placed in a $(2 \ \text{Gpc} / h)^3$ box, every single one of them weighs $3.27 \times 10^8 \ \text{M}_{\odot} / h$. This is comparable to an entire dwarf galaxy and far from resolving star clusters, not to mention individual stars, which would require resorting to a much smaller simulation box. Hydrodynamic simulations, which model the baryonic matter as a separate component typically described by the Euler equations, are computationally more demanding than gravity-only $N$-body simulations and therefore limited to lower resolutions; for instance, the recent \textsc{Flamingo} simulation used $N = 2.8 \times 10^{11}$ particles \citep{Flamingo}. The same holds for ($N$-body only) simulation \textit{suites} such as \textsc{Quijote} ($>$7,000 cosmologies, $N = 1.3 \times 10^8$, \citealt{Quijote}) and even \textsc{AbacusSummit} (97 cosmologies, $N = 3.3 \times 10^{11}$, \citealt{AbacusSummit}) which evolve many different universes with varying cosmological parameters and initial conditions.

Robust, accurate, and precise parameter inference from upcoming galaxy surveys such as \textsc{Euclid} \citep{Euclid} and the Vera C. Rubin Observatory (formerly LSST, \citealt{LSST}) requires, however, the availability of many simulations in large boxes (to mitigate cosmic variance), at high resolution (to exploit small-scale information), and for extensive parameter spaces (to probe $\Lambda$CDM and beyond). Clearly, the computational cost incurred by each of these desiderata calls for a trade-off that depends on the specific application.

Recently, machine learning (ML)---more specifically generative adversarial networks (GANs, \citealt{goodfellow2014generative})---has shown promising results regarding the resolution aspect: given a paired dataset of high-resolution (HR) and low-resolution (LR) simulations, a neural network is trained to augment LR simulations with the missing information contained in their HR counterparts, a task dubbed as \textit{super-resolution} in the ML community. Once trained, the GAN can add the missing HR structure for new LR simulations, without the need for performing an actual HR simulation.

The first super-resolution emulator for cosmological simulations was introduced by \citet{Ramanah2020a}. Their GAN has only $\mathcal{O}(10^4)$ trainable parameters and produces SR samples whose statistics closely match those of the true HR samples. Since their model is directly trained on the density field, however, downstream tasks that require particle positions and/or velocities cannot be carried out. 
A paper series by Yin Li and collaborators \citep{Li_2021} presents a powerful GAN-based super-resolution framework that is trained on displacements fields and velocities \citep{Ni2021}, and can be conditioned on redshift \citep{Zhang2023}. Thus, the output of their GAN can be processed just like a `real' cosmological simulation, i.e.\ haloes can be extracted, merger trees can be built, etc. (cf. \citealt{Schaurecker2021} who build onto the same framework and operate on the density field). An application of their method to fuzzy dark matter simulations has been studied in \citet{Sipp2022}. For completeness, let us mention that GANs have also been used in the context of cosmological simulations for generating entirely new realizations of the cosmic web (i.e.\ without conditioning on LR fields, \citealt{Rodriguez2018, Perraudin2019, Feder2020, Tamosiunas2020, curtis2022cosmic}), for augmenting N-body simulations with baryonic physics \citep{Troster2019, Bernardini2022} or dark sector physics \citep{List2019a}, for mapping dark matter to haloes \citep{ramanah2019painting}, for constructing halo merger trees \citep{robles2022deep}, for encoding simulations in latent space \citep{ullmo2021encoding}, for producing HI maps from hydrodynamic simulations \citep{zamudio2019higan}, and for line intensity mapping (\citealt{Moriwaki_2021}, \citealt{Moriwaki_2021_2}) and 21cm observations \citep{list2020unified, yoshiura2021predicting}.

Clearly, there is an entire \textit{distribution} of possible HR realizations that could belong to each LR field or, in other words, super-resolution is an ill-posed problem. In typical ML applications of super-resolution, the task of the neural network can be viewed as learning a realistic `sharpening' of images by an upscaling factor of e.g.\ 2x, 4x, or 8x. In that case, it is often sufficient if the model produces a single sharp and convincing HR realization given an LR input. For scientific applications, however, it is desirable to have a model capable of producing \textit{diverse} SR realizations for each LR field, i.e.\ sampling from the full distribution $P(\vec{x}_{HR} \mid \vec{x}_{LR})$ and thus allowing for the quantification of the \textit{uncertainties} in the generated small-scale features (both statistically and at the field level). 

Unfortunately, GANs are prone to generate samples from a distribution that is much narrower than the true data distribution, with the most extreme failure case known as \textit{mode collapse}, where all the GAN-produced samples are extremely similar. In particular, when a GAN is conditioned on a highly informative input such as an LR image, the NN has barely any incentive to pay attention to efforts intended to induce stochasticity such as random noise provided as an additional input or dropout and rather focuses on generating a single output consistent with the tight conditional distribution.\footnote{An example for this phenomenon is the GAN-based \textsc{pix2pix} model for the task of image-to-image translation \citep{Isola2017}, where the authors report that noise fed to the generator part of the GAN is simply ignored.} In principle, the framework by \citet{Li_2021} performs a stochastic mapping from each LR input to an HR output, with stochasticity arising from randomly drawn noise that is added after every convolution layer. However, to the best of our knowledge, none of the existing super-resolution models for cosmological simulations has been systematically analyzed regarding their ability to reproduce the true underlying distribution of possible HR fields when conditioned on a specific LR input. 

Although it is possible to carefully design GANs for the task of stochastic super-resolution (see e.g.\ \citealt{leinonen2020stochastic} for an application to atmospheric fields), a more principled approach seems to be switching to an alternative generative model. For instance, normalizing flows (see \citealt{papamakarios2021normalizing} for a review) have been successfully used for stochastic super-resolution \citep{lugmayr2020srflow}. Within the scope of cosmology, \citet{rouhiainen2021normalizing} showed in a proof-of-concept study that normalizing flows are able to generate 2D maps mimicking those extracted from cosmological simulations. In practice, however, their expressiveness is often limited by the restricted way in which they sequentially transform a given base distribution into the data distribution. 

Recently, denoising diffusion models (\citealt{sohldickstein2015deep, ho2020denoising}; hereafter referred to as diffusion models for brevity) have become immensely popular in the ML community as they have been demonstrated to produce images at state-of-the-art quality, outperforming GANs (e.g.\ \citealt{dhariwal2021diffusion}). Apart from their high sample quality, diffusion models have other advantages over GANs such as a likelihood-based, stationary training objective, more stable training, and good coverage of the underlying distribution.
In the context of astronomy, a few applications of diffusion models exist to date, for example for generating galaxy images \citep{smith2022realistic} and other astrophysical images (\citealt{zhao2023can}) and in the context of strong gravitational lensing (\citealt{karchev2022strong}), see also e.g.~\ \citet{adam2022posterior, remy2023probabilistic, legin2024posterior} for cosmology applications that use the closely related score-based generative models.

In this work, we demonstrate for the first time that diffusion models can be harnessed for super-resolving cosmological simulations. We train our model on the displacement fields of two-dimensional cosmological simulations and show that the statistics of the generated SR fields agree well with their HR counterparts. In particular, our model is able to correctly reproduce the diversity of possible HR fields associated with each given LR input. To obtain accurate results down to small scales, we find it crucial to apply a blue filter to the displacement fields before feeding them to the diffusion model, thereby boosting the importance of small scales in the pixel-wise training objective.

This paper is structured as follows. In \autoref{sec:methods}, we summarize the theory of diffusion models, and we discuss our new approach of `filter-boosted training' which is critical to achieve good performance in our case. In \autoref{sec:neural_network}, we discuss all the machine learning technicalities of data generation, network architecture, and training strategy. In \autoref{sec:results}, we then demonstrate the performance of our diffusion super-resolution model in terms of a visual validation of the generated displacement and density fields, as well as in terms of various summary statistics. We also assess in detail, whether the model samples the correct ensemble statistics. Finally, we summarize our results and conclude in \autoref{sec:conclusions}.

\section{Methods}\label{sec:methods}

This section outlines the basis behind diffusion models in the context of super-resolution. Additionally, we introduce the motivation and functionality of filter-boosted training.

\subsection{Denoising Diffusion Models}\label{sec:diffusion_model}

Diffusion models (\citealt{sohldickstein2015deep}, \citealt{ho2020denoising}) try to learn to reverse a destructive diffusion process in which increasingly more noise is added to an image $\vec{x}_0 \sim q(\vec{x}_0)$. The noising process is implemented as a Markov chain of $T$ steps, adding Gaussian noise in each step, which is represented by the \textit{forward process} $q(\vec{x}_{t} \mid \vec{x}_{t-1})$,

\begin{equation}
  \begin{array}{l}
    q(\vec{x}_t \mid \vec{x}_{t-1})=\mathcal{N}\left(\vec{x}_t ; \sqrt{1-\beta_t} \vec{x}_{t-1}, \beta_t \mathbf{I}\right), \\ 
    q(\vec{x}_{1: T} \mid \vec{x}_0)=\prod_{t=1}^T q(\vec{x}_t \mid \vec{x}_{t-1}),
  \end{array}
\end{equation}
where $\mathcal{N}(\bm{\mu}, \boldsymbol{\Sigma})$
is the normal distribution with mean $\bm{\mu}$ and covariance matrix $\boldsymbol{\Sigma}$.

The “amount” of noise is fixed by a variance schedule $\left\{\beta_t \in(0,1)\right\}_{t=1}^T$, where $\beta_1<\beta_2<\cdots<\beta_T$. The values of $\beta$ must be small in comparison to the (normalized) images $\vec{x}_0$. Due to the Gaussian property of the noise, one can directly sample $\vec{x}_t$ at an arbitrary time step $t$, without the need to compute all the $t-1$ previous noisy images before, as

\begin{equation}
     \vec{x}_t = \sqrt{\bar{\alpha}_t}\vec{x}_{0} + \sqrt{1 - \bar{\alpha}_t} \boldsymbol{\epsilon},
\end{equation}

\noindent using $\alpha_t=1-\beta_t \text { with } \bar{\alpha}_t=\prod_{i=1}^t \alpha_i$ and $\boldsymbol{\epsilon} \sim \mathcal{N}(\vec{0}, \mathbf{I})$. 
The inverse of $q(\vec{x}_{t} \mid \vec{x}_{t-1})$, i.e.\,\,$q(\vec{x}_{t-1} \mid \vec{x}_{t})$, called the \textit{backward process} \citep[with the same functional form as the forward process, ][]{feller1949}, is an unknown.  Due to the destructive nature of the process, the inversion is not possible without knowledge of the underlying data distribution (there are many more trajectories from the data manifold to noisy data than from noisy data onto the data manifold). Therefore, in diffusion models, one utilizes a neural network (NN), with learnable parameters $\boldsymbol{\theta}$ to learn the reverse process $p_{\boldsymbol{\theta}}(\vec{x}_{t-1} \mid \vec{x}_{t})$,
\begin{subequations}
\begin{align}
    p_{\boldsymbol{\theta}}(\vec{x}_{t-1} \mid \vec{x}_{t}) & =\mathcal{N}(\vec{x}_{t-1} ; \bm{\mu}_{\boldsymbol{\theta}}(\vec{x}_{t}, t), \boldsymbol{\Sigma}_{\boldsymbol{\theta}}(\vec{x}_{t}, t)), \label{eq:reverse_process}
    \intertext{hence}
    p_{\boldsymbol{\theta}}(\vec{x}_{0: T}) &= \prod_{t=1}^T p_{\boldsymbol{\theta}}(\vec{x}_{t-1} \mid \vec{x}_{t}).
\end{align}
\end{subequations}
Contrary to expectations, \cite{ho2020denoising} found that setting the variance to a fixed value, i.e.\ $\boldsymbol{\Sigma}_{\boldsymbol{\theta}}(\vec{x}_{t}, t) = \sigma_t^2 \mathbf{I}$\, produces similar results compared to trainable variances; hence, we set $\sigma_t^2 = \beta_t$ for all tested cases. 
Also, they proposed a simplified objective loss function $L_t$ for training the NN: 
\begin{align} \label{eq:simple_loss}
    L_t & =  \mathbb{E}_{t\sim[1,T],\vec{x}_0,\boldsymbol{\epsilon}}\bigg[\|\boldsymbol{\epsilon}-\boldsymbol{\epsilon}_{\boldsymbol{\theta}}(\vec{x}_t,t)\|^2\bigg],
\end{align}
where $\boldsymbol{\epsilon}_{\boldsymbol{\theta}}(\vec{x}_t,t)$ is the NN estimate of the noise $\boldsymbol{\epsilon}$ present in the image $\vec{x}_t$, and $t\sim[1,T]$ denotes a discrete uniform draw of the step $t$.

The aim of inference is to generate $\vec{x}_0 \sim p(\vec{x}_0)$ based on pure noise. In other words, one wants to sample $\vec{x}_{t-1}$ given a known $\vec{x}_{t}$ traversing the Markov chain until $t=0$. Sampling $\vec{x}_{t-1} \sim p_{\boldsymbol{\theta}}(\vec{x}_{t} \mid \vec{x}_{t-1})$ requires computing
\begin{equation}
    \vec{x} _{t-1}=\frac{1}{\sqrt{\alpha_{t}}}\left(\vec{x}_{t}-\frac{\beta_{t}}{\sqrt{1-\bar{\alpha}_{t}}}\boldsymbol{\epsilon}_{\boldsymbol{\theta}}(\vec{x}_{t},t)\right)+\sigma_{t}\vec{z}, \label{eq:sampling}
\end{equation}
\noindent where $\vec{z}\sim\mathcal{N}(\vec{0},\mathbf{I})$. Here, the first term is the mean estimate $\bm{\mu}_{\boldsymbol{\theta}}(\vec{x}_{t}, t)$ in Eq.~\eqref{eq:reverse_process} provided by the NN, which is perturbed by Gaussian noise with standard deviation $\sigma_t$, resembling a Langevin sampling step \citep{welling2011bayesian}.

\subsubsection{Diffusion Models For Super-Resolution}

To condition diffusion models for super-resolution tasks, for an HR image $\vec{x}_0$ with a corresponding LR image $\vec{y}_0$, we require the NN to learn the conditional distribution $p_{\boldsymbol{\theta}}(\vec{x}_0 \mid \vec{y}_0)$. For this, we provide it with the additional information of $\vec{y}_0$; implying that the mean in Eq.~\eqref{eq:reverse_process} and the predicted noise in the loss function in Eq.~\eqref{eq:simple_loss} become $\bm{\mu}_{\boldsymbol{\theta}}(\vec{x}_t, \vec{y}_0, t)$ and $\boldsymbol{\epsilon}_{\boldsymbol{\theta}}(\vec{x}_t, \vec{y}_0, t)$, respectively.
The conditioning on the LR image is realized by a channel-wise concatenation of the LR image $\vec{y}_0$ with the reverse process input $\vec{x}_t$ at each time step of the Markov chain, as proposed by \cite{saharia2021superres} and \cite{ho2021cascaded}.

\subsection{Diffusion models for large-scale structure}
In a cosmological setting, we use diffusion models to increase the resolution on small scales where we know that in principle structure formation ends in a very non-linear but well-defined high-entropy state: the density profiles of haloes are well described by profiles with few degrees of freedom \citep{1965TrAlm...5...87E, Navarro1997}, which are at first order independent of larger scales. One can proceed in different ways, by either directly increasing the resolution of the density distribution (or some observable field) in Eulerian space, or by modeling the Lagrangian map between the initial (unperturbed) position of a fluid element $\vec{q}$ and its later (evolved) Eulerian space position $\vec{X}$
\begin{align}
    \vec{q}\mapsto \vec{X}(\vec{q},a) = \vec{q} + \vec{\mPsi}(\vec{q},a),
\end{align}
from which other statistics can be derived. In particular, the Eulerian density field $\delta(\vec{X})$ is given by the projection

\begin{align}
    1+\delta(\vec{X}) = \int \dd^2 q\;\delta_{\text{D}}(\vec{X}-\vec{q}-\vec{\mPsi})\;, \label{eq:density_eulerian}
\end{align}

\noindent where $\delta_{\text{D}}(\cdot)$ is the Dirac-$\delta$ distribution. It is a well known fact that the density field described by Eq.~\eqref{eq:density_eulerian} has low regularity (and in principle infinite density caustics), while the displacement field is significantly more regular. In addition, the mapping from displacements to densities is bijective only in the linear regime. At later times, multiple fluid streams can occupy the same Eulerian space position $\vec{X}$, and reconstructing the displacement field $\vec{\mPsi}$ from a given density field $\delta$ is no longer possible uniquely (e.g.\ \citealt{feng2018exploring}). It is thus not surprising that \cite{Li_2021} found that training their GAN model on the displacement $\vec{\mPsi}$ instead of the density field resulted in much better outcomes, and we will proceed similarly here by considering super-resolution for the displacement field.
In future work, we will extend our framework to also super-resolve the velocities, which can simply be treated as additional channels (similarly as for the super-resolution GAN by \citealt{Ni2021}).

\subsection{Filter-boosted Training }\label{sec:filter_training}

While \cite{Li_2021} trained their GAN super-resolution model on the displacement field, they additionally provided their discriminator network the
density fields computed directly from the displacement fields produced by the generator network. They report that this not only greatly enhanced the accuracy at small scales of the matter power spectra, but also the visual sharpness of the images. However, such an approach is not applicable to diffusion models, given that the loss function is a direct comparison of predicted and actual noise (see Eq.~\ref{eq:simple_loss}), so that such additional supplemental fields derived from the predicted displacements are not readily integrated in the process.

During our early implementation phase, we found that the pixel-wise loss was dominated by the information from large scales, and our NN was therefore not able to accurately reproduce the matter power spectra at small scales. To overcome this limitation and increase the importance of these scales, we propose a novel training technique. By applying a (problem-dependent) `blue' filter $f$ to the training data, it is possible to increase the amplitude of small scales relative to large scales, thus altering their relative contribution to the pixel-wise loss. We call this technique `\textit{Filter-boosted Training}'. 

\begin{figure}[!t]
  \centering
  \includegraphics[width=1.\linewidth]{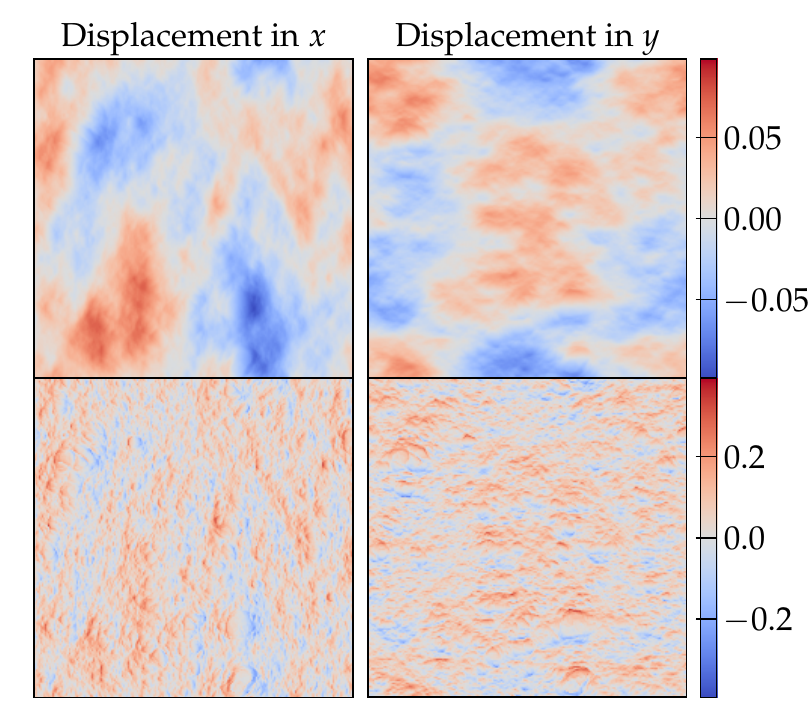}
  \caption[]{Comparison between the original displacement field $\vec{\mPsi} (\textit{top})$ and filtered displacement field $f(\vec{\mPsi})$ (\textit{bottom}) using $\gamma = 1$. Directly evident are much more pronounced small-scale features in the filtered displacement field. The diffusion model is therefore forced to gain a good understanding of these small-scale features in order to accurately estimate the noise added to the filtered fields in the forward process.}
  \label{im:transform}
\end{figure}

In the context of super-resolving cosmological simulations, we apply the following generalized Laplacian operator to the displacement fields,

\begin{equation} \label{eq:filter}
    \vec{\mPsi}^\prime = f(\vec{\mPsi}) = \mathcal{F}^{-1}(\mathcal{F}(\vec{\mPsi}) \norm{\vec{k}}^\gamma),
\end{equation}

\noindent where $\vec{k}$ is the wave vector and $\mathcal{F}$ refers to the (Discrete) Fourier Transform, here understood to be separately applied to the $x$- and $y$- components of the displacement field. Excluding the pathological case $\|\vec{k}\| = 0$ (by leaving the `DC mode' unchanged when applying the filter), this mapping is invertible, which allows us to apply the inverse filter to the outputs of the diffusion model to obtain displacement fields. Filter coefficients $\gamma > 0$ ($\gamma < 0$) increase the amplitude of small (large) scales in the data. An example of filtered displacement fields for $\gamma = 1$ can be found in Figure~\ref{im:transform}.

In principle, much more general filters can be envisioned, or could even be learned from data as part of a hyperparameter tuning; we were guided by simplicity and the goal to achieve a pixel-wise loss that has a power spectrum (and thus `roughness') similar to that of the density field. We will discuss this in more detail below. 

\section{Neural Network: Architecture and Training}\label{sec:neural_network}

This section presents the architecture of the utilized NN and details the training data generation as well as training specifics.

\subsection{Data Generation}\label{sec:data_generation}

Our training data in this study comes from a suite of fast approximate $N$-body simulations with varying initial conditions (i.e.\ different random seeds for the perturbations). Specifically, we train on the particle displacements at final time in respect to the initial conditions at time zero on an unperturbed grid. 
Further, since this is a proof-of-concept study, we use 2D simulations to reduce the computational complexity of the data generation process, as well as that of training and sampling with the diffusion model.

Our fast simulations perform eight time steps with the \textsc{LPTFrog} time integrator recently introduced by \citet{list2023perturbationtheory}, which is consistent with the Zel'dovich approximation on large scales (similar to \textsc{FastPM}, \citealt{FastPM}) and therefore requires much fewer steps to accurately reproduce the correct growth on large scales in comparison with a standard symplectic leapfrog integrator. For computing the accelerations, we use the particle mesh method, solving the Poisson equation with an exact (spectral) Laplacian kernel (i.e.\ $-k^{2}$) and a fourth-order finite-difference gradient kernel.

We consider the evolution of a two-dimensional matter distribution (in a three-dimensional Universe with standard gravity), dominated by non-relativistic collisionless matter (i.e.\ Einstein-de~Sitter) with a power-law (scale-free) initial perturbation spectrum
\begin{align}
    P_0(k) = A\;k^{n_s}\;,\quad\text{where here $n_s=-1.5$}\;.
\end{align}
We fix the cosmology throughout the paper and leave an extension to varying cosmologies to future work.

The choice of power-law perturbations is motivated by the recent renewed interest in scale-free simulations to study the detailed performance of structure formation models (e.g.\ \citealt{Joyce_2020}). 
In 3D, the primordial power spectrum that roughly scales as $P_{\mathrm{prim}}(k) \propto k$ \citep{aghanim2020planck} changes its slope to $\propto k^{-3}$ on scales $k \gg k_{\mathrm{eq}}$ during the radiation-dominated epoch, where $k_{\mathrm{eq}}$ denotes the scale that enters the Hubble horizon at the time of matter-radiation equality. Thus, this CDM small-scale limit value of $n_s = -3$ in 3D would correspond to a choice of $n_s = -2$ in 2D, and our value of $n_s = -1.5$ represents a slightly shallower spectrum (which avoids the issues arising for $n_s \to -2$ such as an infinite standard deviation of the density field due to infrared divergence and the almost instantaneous collapse of structures on vastly different scales in view of the dimensionless power spectrum $\Delta^2(k) \propto k^2 \, P_0(k)$ losing its scale dependence).
Due to the self-similar nature of structure formation for a scale-free initial power spectrum and EdS cosmology, super-resolution should be possible across all scales and times -- an aspect that we will investigate in future work.

The box size and initial scale factor $a_{\text{ini}}$ were set arbitrarily to 1 as the single relevant scale in scale-free universes is the non-linear scale. Therefore, the scale factor $a$ has no physical meaning and should only be interpreted as a timescale. We chose the end time as $a_{\text{end}} = 15$, by which point non-linear structures have formed.

For the LR and HR simulations, we used $64 \times 64$ and $256 \times 256$ $N$-body particles, respectively, which we placed onto a regular two-dimensional lattice in Lagrangian space at time zero and then displaced using second-order Lagrangian perturbation theory (2LPT) to obtain the initial conditions for the $N$-body simulations at the initial time $a_{\text{ini}}$.
The outcome of each $N$-body simulation is a 2-channel `image' of resolution $64 \times 64$ (LR) and $256 \times 256$ (HR), where the two values in each pixel are given by the $x-$ and $y-$components of the total displacement (2LPT up to $a_{\text{ini}}$ + simulation up to $a_{\text{end}}$) of the $N$-body particle that started its trajectory in that pixel at time zero. We then transformed the pixel values via Eq.~\eqref{eq:filter}, and subsequently normalized them between $[-1, 1]$ based on the ensemble maximum and minimum.

\begin{figure*}[!ht]
  \centering
  \includegraphics[width=1\linewidth]{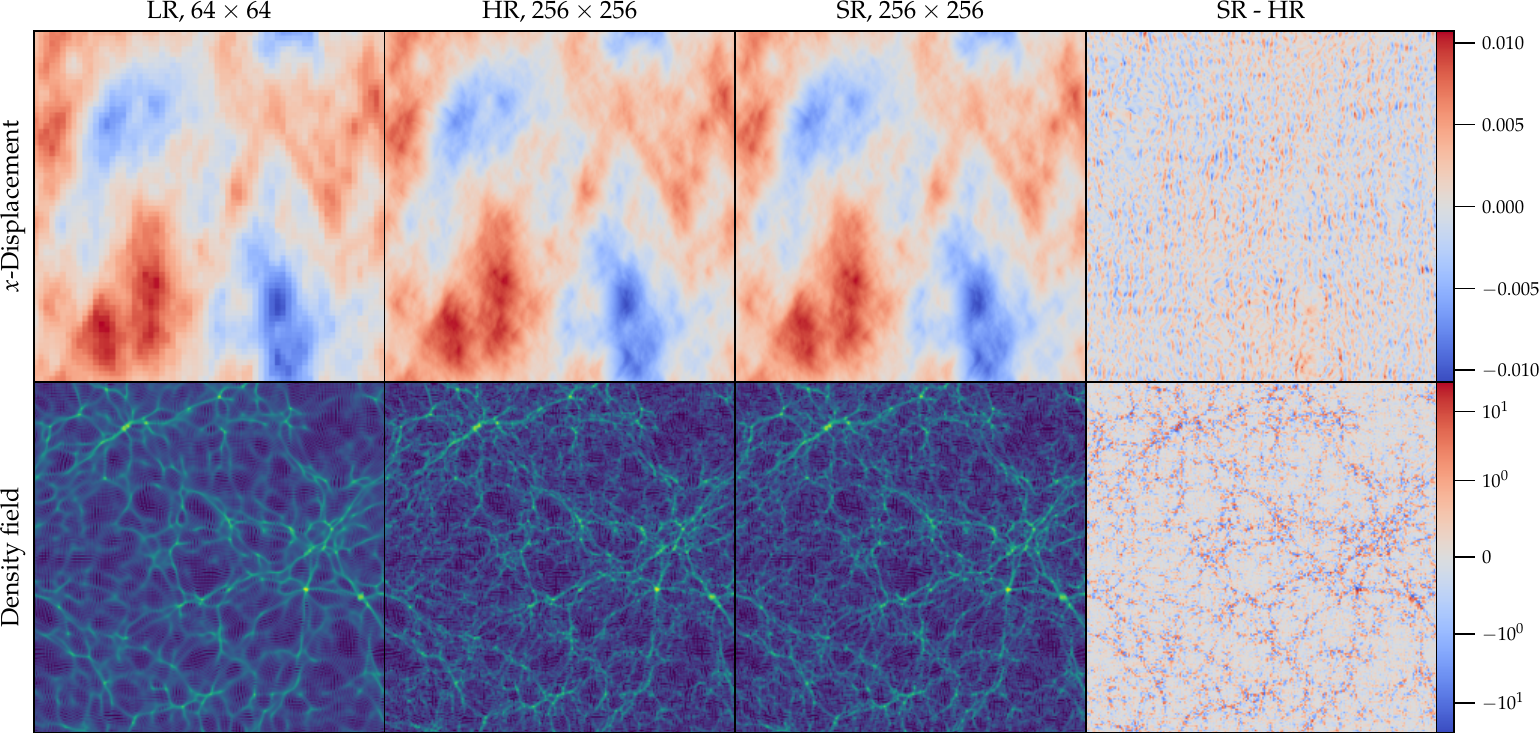}
  \caption[]{Visual comparison between a randomly selected LR simulation, its HR counterpart, and a SR sample generated by our model (trained on filtered data using $\gamma = 1$). The top row represents the $x$-displacement $\bm{\mPsi}_x$, and the bottom row the density field.
  The rightmost column shows the residuals between SR and HR (that indicate the difference between two possible realisations of small scales given the LR constraint). The SR density field is computed from the output of our model by using cloud-in-cell interpolation. In order to increase the contrast of the density fields, we plot $\log(1 + \delta + \epsilon)$ with $\epsilon = 0.2$. The LR density is computed from the Fourier-upscaled LR displacement field (matching the resolution of the HR simulation).} 
  \label{im:comp}
\end{figure*}

\subsection{Model Architecture}\label{sec:architecture}

The currently most effective architecture for diffusion models is given by so-called U-nets \citep{ronneberger2015unet}. U-nets are comprised of the encoder part, which downsamples the input image towards the bottleneck, and the decoder part, which in turn decodes the information from the bottleneck and upsamples towards the original input resolution. At each resolution level, information is passed from the encoder to the decoder part via a `skip-connection'.
As a result of this architecture, the NN is forced to identify the key underlying patterns and to pass only the most relevant information through the bottleneck. Our architecture is driven by simplicity and based on the NN employed in \cite{ho2021cascaded}, starting from the input resolution, downsampling until a resolution of $8 \times 8$ is reached, and upsampling again to the full resolution.
Each block in the encoding path of the U-net consists of two residual convolutional layers and one convolutional layer, which halves the spatial resolution. For all convolutions, we use a kernel size of $3 \times 3$ with periodic padding in order to account for the periodicity of the simulations. The blocks in the decoding path consist of exactly the same components, however, the last transposed convolutional layer in each block doubles the spatial resolution. We utilize the GELU (Gaussian Error Linear Unit; \citealt{GELU}) as the activation function in-between each convolution in the encoder, bottleneck, and decoder. We condition our model on the current diffusion step based on sinusoidal embedding.

In order for the spatial resolution of the LR displacement field $\vec{y}_0$ to match that of the HR image $\vec{y}_t$ so that the two can be concatenated along the channel axis and provided as an input to the U-net (e.g. \citealt{saharia2021superres}, \citealt{ho2021cascaded}), we upsample $\vec{y}_0$ using Fourier interpolation. Specifically, we zero-pad the Fourier-transformed LR field $\vec{y}_0$ to high resolution and transform the result back to the spatial domain (see e.g.~\ \citealt{Hahn_Angulo_2016} for interpolation of the displacement field, which exploits the manifold structure of the phase-space density for a cold medium). This procedure leaves the phases and amplitudes of all modes contained in the LR field unaffected, and there is no power on scales beyond the LR Nyquist frequency. Given this input, the task of our super-resolution model is then to add new structure at those small scales and to consistently mimic the backreaction onto larger scales.

We rescale both skip-connections and residual connections by $1/\sqrt{2}$, which was shown to slightly improve outcome results in some cases (e.g. \citealt{dhariwal2021diffusion}). We further implement self-attention layers \citep{vaswani2017attention}, appended after the last transposed convolution at resolution $8\times8$ and $16\times16$. In total, our model has 42 million trainable parameters. 
During inference, we provide the noise latent variable $\vec{z}$ to the NN, together with an LR image $\vec{y}_0$. Let us emphasize that apart from the use of Fourier upscaling for the conditioning on the LR images instead of the commonly employed bilinear interpolation, no `cosmology-specific' ideas or knowledge have been built into our model, and it has simply been trained to perform the task of {\it image} super-resolution, where the images are given by displacement fields in our case.

\subsection{Training Specifics}\label{sec:training}

We set $T = 2{,}000$, $\beta_{1} = 10^{-4} / 2$ and $\beta_{T} = 0.02 / 2$ for all experiments, and opted for a linear schedule for the $\beta$-values. To identify optimal settings, we swept over different dropout rates in the encoder part, the number of attention heads at resolution $8 \times 8$ and $16 \times 16$, and different learning rates for the Adam optimizer (we utilized no warm-up and default values, \citealt{kingma2017adam}). We further implemented \textit{conditioning augmentation} in the form of \textit{blurring augmentation} as reported in the super-resolution work of \cite{ho2021cascaded} to be a crucial implementation to generate visually sharp images, and tested different  standard deviations for a $3 \times 3$ Gaussian kernel. 

We found that no dropout, two attention heads, a learning rate of $2 \times 10^{-4}$ and a standard deviation of the Gaussian kernel randomly sampled from the fixed range of (0.4, 0.6) during training to be optimal. We did not test different batch sizes and simply chose the highest batch size possible given our setup, which was 50. Throughout the training, we applied blurring augmentation as well as a horizontal and vertical flip to the training data, each with a 50\% probability. Furthermore, we applied an exponential moving average (EMA) with an EMA rate of 0.9999 to the weights of the NN.

We leave further optimizations used in state-of-the-art diffusion models (e.g. \citealt{dhariwal2021diffusion}) such as a trainable variance (which would enable sampling with fewer diffusion steps during inference) and a `hybrid' loss objective \citep{nichol2021improved} to future work. We performed the training on a single NVIDIA A100, which took three days. 

\section{Results}\label{sec:results}

We start our analysis with a visual inspection of a single HR/SR pair and then proceed to validate our model quantitatively in terms of summary statistics and assess if our model produces correct ensemble statistics. As we found a filter coefficient of $\gamma = 1$ to perform best (see below), all results use this value unless otherwise stated.

\subsection{Visual analysis and comparison}
As a first validation, we visually analyze the difference between a single HR and SR simulation. Figure~\ref{im:comp} presents a comparison between the $x$-displacement (\textit{top row}) and density field (\textit{bottom row}) generated from an LR simulation taken from our test set. 
For comparability, the LR density is computed from the Fourier-interpolated LR displacement field, representing a `trivially super-resolved' density field without any structures above the LR Nyquist scale. 

We only show the $x$-displacement, as the $y$-displacement would provide qualitatively similar information. Both the displacement and the density fields demonstrate that our model generates convincing SR simulations using filter-boosted training, which are visually indistinguishable from the HR simulation. This is further supported by the small residuals of the $x$-displacements. A comparison between the LR, HR, and SR simulations demonstrates that the NN is successful in authentically generating information on the smallest scales (even reproducing simulation artifacts due to particle noise). 

\begin{figure}[!t]
  \centering
  \includegraphics[width=0.8\linewidth]{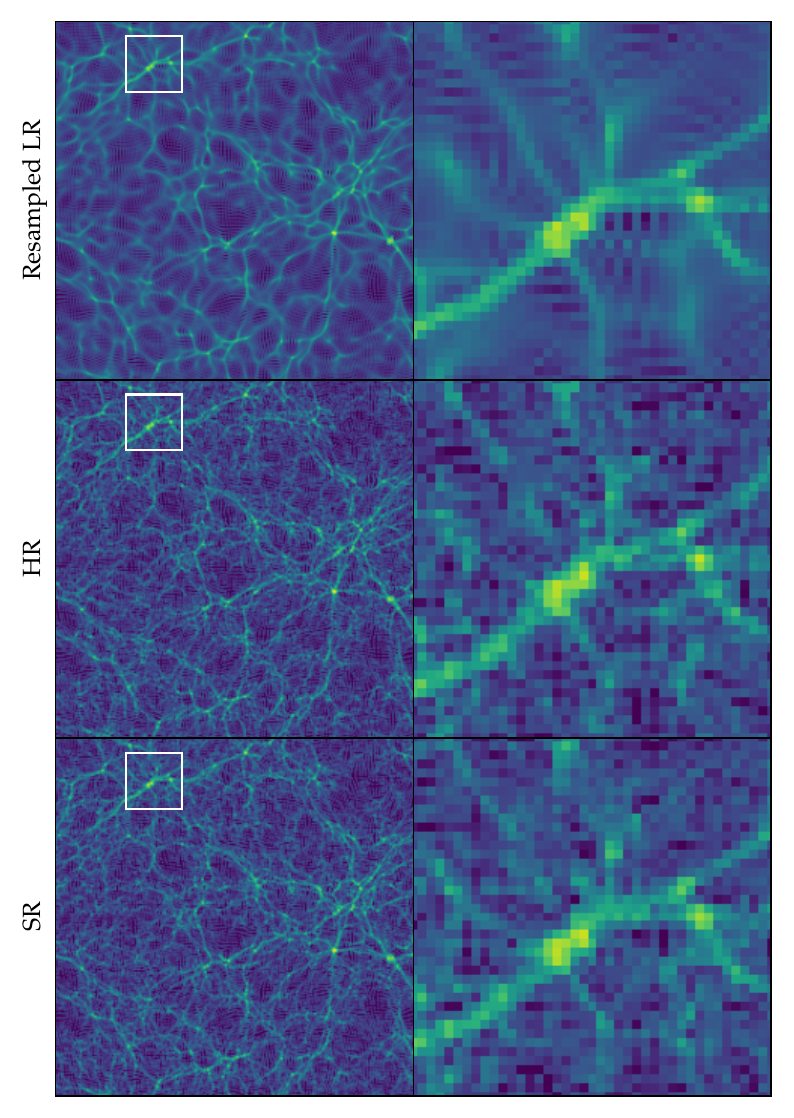}
  \caption[]{Zoom into the white box of the density fields presented in Figure~\ref{im:comp}. The top row represents the density field from the LR simulation (upsampled with Fourier interpolation to match the resolution of the HR simulation), the middle row from the HR simulation and bottom row from the SR equivalent. The inset showcases that the SR authentically reproduces the high-density nodes and even discreteness artifacts of the individual $N$-body particles.}
  \label{im:zoom}
\end{figure}

We inspect the small-scale differences of the density fields presented in Figure~\ref{im:comp} in Figure~\ref{im:zoom}. Here we zoom into an elongated filamentary structure inside the white box. Clearly visible are small-scale structures in the zoomed inset from the HR and SR simulations, while the LR simulation (upsampled with Fourier interpolation to match the resolution of the HR simulation) only contains the most pronounced filaments. Going from LR to HR, high-density nodes emerge, which are reproduced authentically by the SR realization.

\subsection{Statistics for one LR image \texorpdfstring{$\mapsto$}{to} many SR images}

\begin{figure}[!t]
  \centering
  \includegraphics[width=1\linewidth]{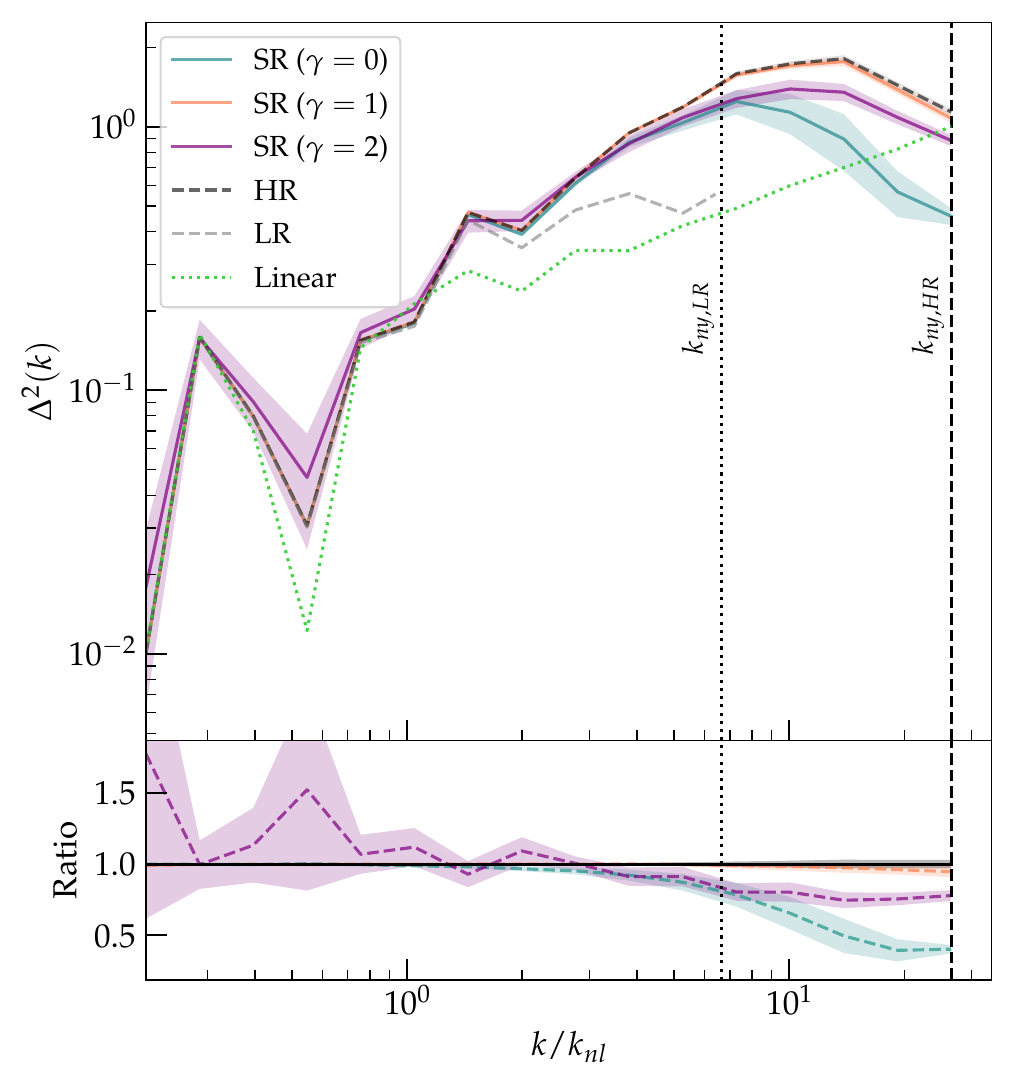}
  \caption[]{\textit{Top panel:} Dimensionless power spectrum $\Delta^2$ of the SR based on one LR/HR pair for different filters. We tested applying no filter ($\gamma = 0$), a generalized Laplacian ($\gamma = 1$), and Laplacian operator ($\gamma = 2$). The original LR and HR power spectra are represented by the \textit{gray dashed} and \textit{black dashed} lines, respectively. The vertical lines show the Nyquist wave numbers. The shaded areas represent the 1$\sigma$ error bars generated from 100 SR and HR simulations based on one single LR simulation. \textit{Bottom panel:} Ratio between all respective power spectra in comparison to the HR simulation (= unity).}
  \label{im:transform_power}
\end{figure}

We now study the impact of the filter on the quality of the NN-generated SR images. Specifically, we analyze the matter power spectrum $P(k)$, which quantifies variations of the density contrast as a function of scale, and encompasses the \textit{full} summary information for a \textit{Gaussian} random field. 

We consider the cases $\gamma = 0$ (i.e.\ no filter), $\gamma = 1$ (square-root of the negative Laplacian operator $(-\nabla^2)^{\nicefrac{1}{2}}$, see e.g.\ \citealt{kwasnicki2017ten}), and $\gamma = 2$ (negative Laplacian $-\nabla^2$). Figure~\ref{im:transform_power} shows the dimensionless power spectrum \mbox{$\Delta^2(k) \equiv k^2P(k) / 2 \pi$}, where $k = \norm{\vec{k}}$, for all tested filters based on the LR/HR pair from Figure~\ref{im:comp}. The scale $k_{nl}$ defines the transition from the linear ($k < k_{nl}$) to the non-linear ($k \geq k_{nl}$) regime. Considering a 2D universe, we define the non-linear wave number $k_{nl}$ via,

\begin{equation}
    2\pi \int_{0}^{k_{nl}} \mathrm{d}k \, k \, P_{\text{lin}}(k, a) = 1,
\end{equation}

\noindent yielding $k_{nl} = 1 / (4 \pi A a^2)^2$ through $P_{\text{lin}}(k, a) = Aa^2 k^{-3/2}$. The vertical lines indicate the Nyquist wave numbers \mbox{$k_{ny} = \pi N_{\text{mesh}}/L_{\text{box}}$} of the LR and HR simulations, with $N_{\text{mesh}}$ being the number of grid cells per dimension and $L_{\text{box}}$ the length of the simulation box. The shaded areas indicate the $1\sigma$ scatter of 100 SR simulations generated from a single LR simulation. On the bottom panel, we present the ratios between the power spectrum of each HR/SR realization and the HR ensemble mean, for 100 HR and SR simulations.

Clearly, filter-boosted training with $\gamma = 1$ outperforms the other filters on all scales: the SR mean power spectrum deviates from the HR ensemble mean by less than 2\% for $k < k_{ny, LR}$ and less than 5\% down to $k_{ny, HR}$.
Applying no filter ($\gamma = 0$) performs the worst on the smallest scales, where we observe a maximum difference of more than 50$\%$, but correctly reproduces the power spectrum on the largest scales. 
A value of $\gamma = 2$ performs somewhat better on small scales, however, leads to bias and a large scatter on large scales. 

A key contribution of this work is our model's ability to perform \textit{stochastic} super-resolution. In other words, for a given LR image, our model is able to generate varied HR realizations consistent with the features imposed by the LR image. While some existing (GAN-based) cosmological super-resolution models are deterministic (e.g. \citealt{Ramanah2020a}), others such as the model presented in \citet[which is based on \textsc{StyleGAN2}, \citealt{karras2019stylebased}, \citealt{karras2020analyzing}]{Li_2021} in principle contain random components, e.g. noise added in several layers of the NN. However, a systematic demonstration of consistency between the SR and HR distributions, conditioned on an LR simulation, has (to the best of our knowledge) not yet been presented in the literature. 

\begin{figure}[!t]
  \centering
  \includegraphics[width=1\linewidth]{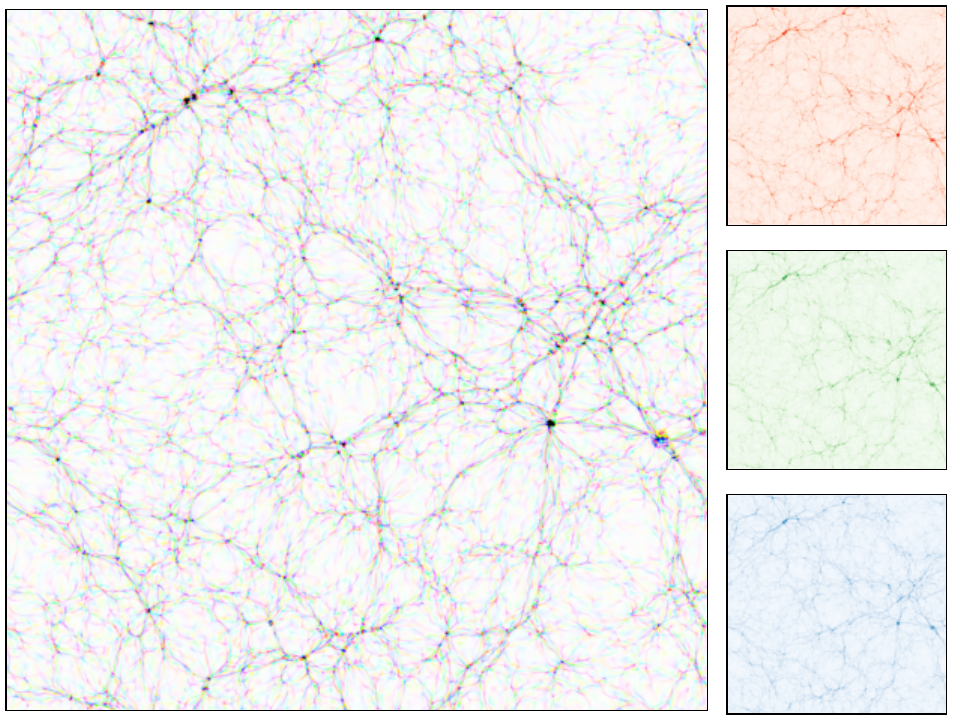}
  \caption[]{RGB composite image (\textit{left column}) generated from 3 different SR simulations (\textit{second column}) that belong to the same LR image. Each SR simulation is represented by one color channel (red, green, and blue). We upsampled the image via Fourier interpolation to a resolution of 1,024$\times$1,024 in order to suppress particle noise and highlight the structural differences between the realizations. When plotted side by side, identifying the differences between the SR images is challenging;
  however, the composite image highlights subtle differences along the filamentary structures and at various high density nodes.}
  \label{im:rgb}
\end{figure}

From a technical point of view, the stochasticity in our model predictions arises from the fact that the reverse process is initialized with a random Gaussian noise realization $\vec{x}_T$, which is then iteratively `denoised' (conditioned on the given LR image $\vec{y}_0$), i.e. brought closer to a point on the data manifold. Moreover, the trajectory from the initial random noise $\vec{x}_T$ to the generated output $\vec{x}_0$ is also stochastic, as noise is added in each of the $T$ iterative steps performed during sampling (see the last term in Eq.~\ref{eq:sampling}). 

To provide some intuition for the qualitative differences between different SR realizations of our model that belong to the same LR simulation, we present a color composite of three different SR simulations based on the same initial LR simulation in Figure~\ref{im:rgb}. Gray-scale features (such as white voids and black clusters) are shared by all three SR realizations, whereas color features are exclusive to one (or two) of them.

While the most prominent nodes are present in all three SR simulations, the filamentary structure, as well as the locations of smaller nodes, are clearly distinct.

\begin{figure}[!t]
  \centering
  \includegraphics[width=1\linewidth]{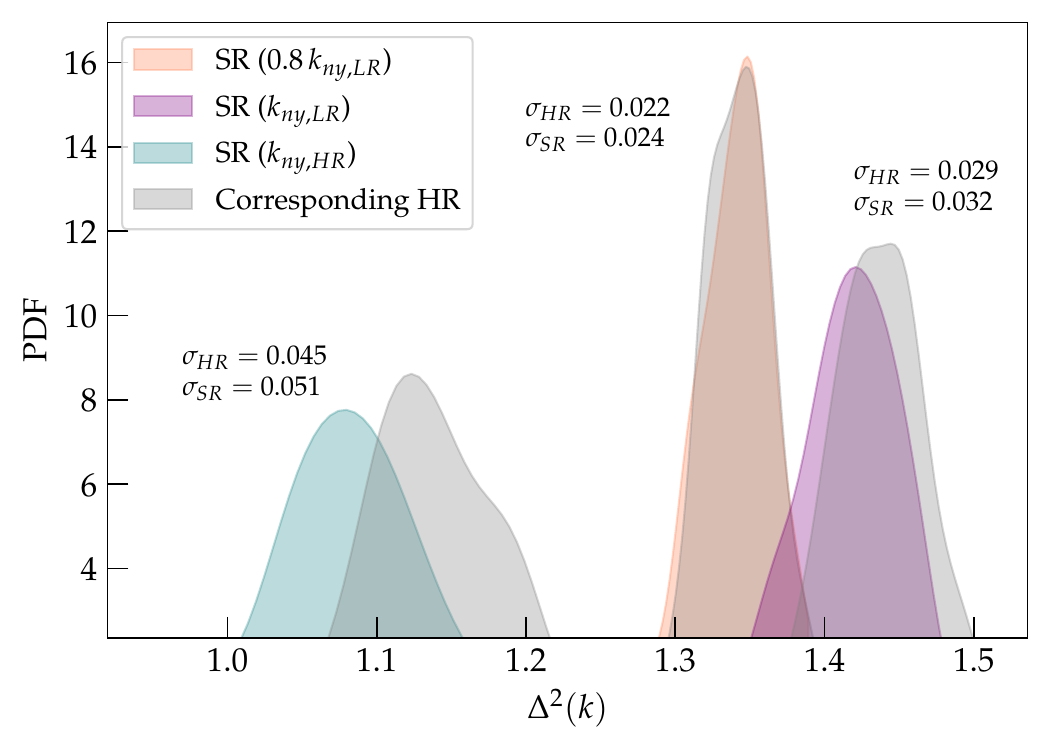}
  \caption[]{PDF of the HR and SR ($\gamma = 1$) dimensionless power spectra from Figure~\ref{im:transform_power} evaluated based on the slices along the $0.8\; k_{ny,LR}$, $k_{ny,LR}$ and $k_{ny,HR}$ bins. The standard deviations calculated for the respective slice of each HR and SR pair are shown directly next to each pair.}
  \label{im:sigma_distribution}
\end{figure}

To investigate if our model correctly reproduces the distribution of HR simulations conditioned on the same LR simulation,  
we inspect the distribution of the dimensionless power spectrum $\Delta^2(k)$ in Figure~\ref{im:transform_power} for three vertical slices at $k = 0.8 \ k_{ny,LR}$, $k_{ny,LR}$, and $k_{ny,HR}$. Specifically, we compute kernel density estimates based on 100 HR and SR simulations conditioned on the LR simulation shown in Figures~\ref{im:comp} and \ref{im:zoom}, and plot the results in Figure~\ref{im:sigma_distribution}.

As expected, the width of the PDF increases for smaller scales as these features are less constrained by the LR simulation. In fact, notice that without any backreaction from small scales beyond the LR simulation's Nyquist frequency, the HR distribution of $\Delta^2(k)$ for $k \leq k_{ny, LR}$ would be a Dirac $\delta$-distribution (recall that we condition on a single LR simulation). Thus, our model is required to not only add new structure on scales $k > k_{ny, LR}$, but to also adjust features contained in the LR simulation to such a degree as to be consistent with the backreaction from smaller to larger scales caused by the non-linear structure growth.

\begin{figure}[!hb]
  \centering
  \includegraphics[width=0.95\linewidth]{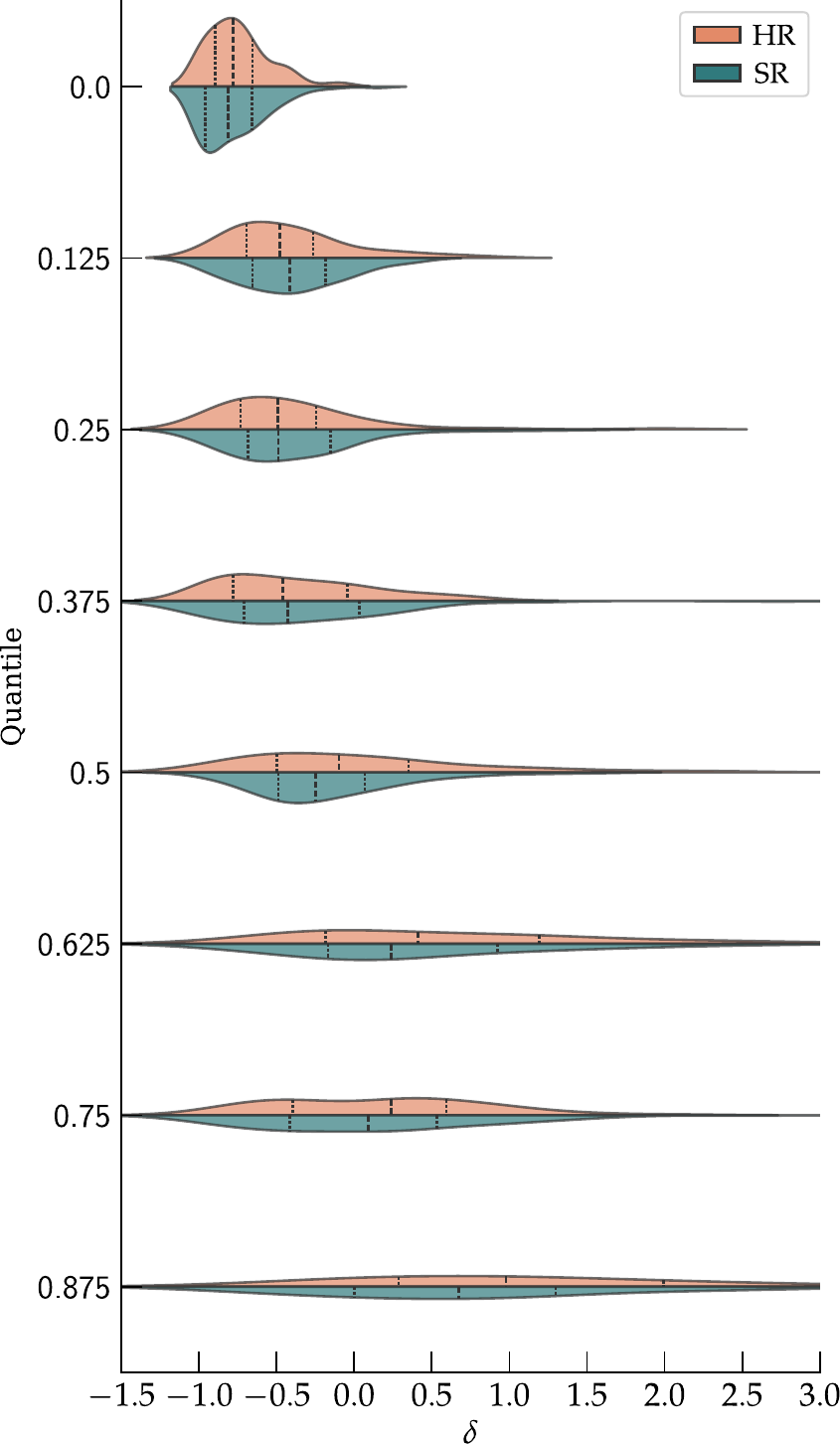}
  \caption[]{Comparison of the HR/SR distributions for the density contrast $\delta$ conditioned on the same LR realization. The rows correspond to eight pixels which, when going from top to bottom, reside in regions of gradually increasing density (as indicated by the corresponding quantile w.r.t.\ the LR density contrast, see main text).  
  For each pixel, the HR and SR distributions have been computed over 100 realizations. The dashed / dotted vertical lines show the quartiles. Supports extending below $\delta = -1$ are artifacts due to the finite band-width of the Gaussian kernel density estimator.
  The pixel distributions between HR and SR match very well across all density regions.
}
  \label{im:violin}
\end{figure}

\begin{figure}[!ht]
  \centering
  \includegraphics[width=1\linewidth]{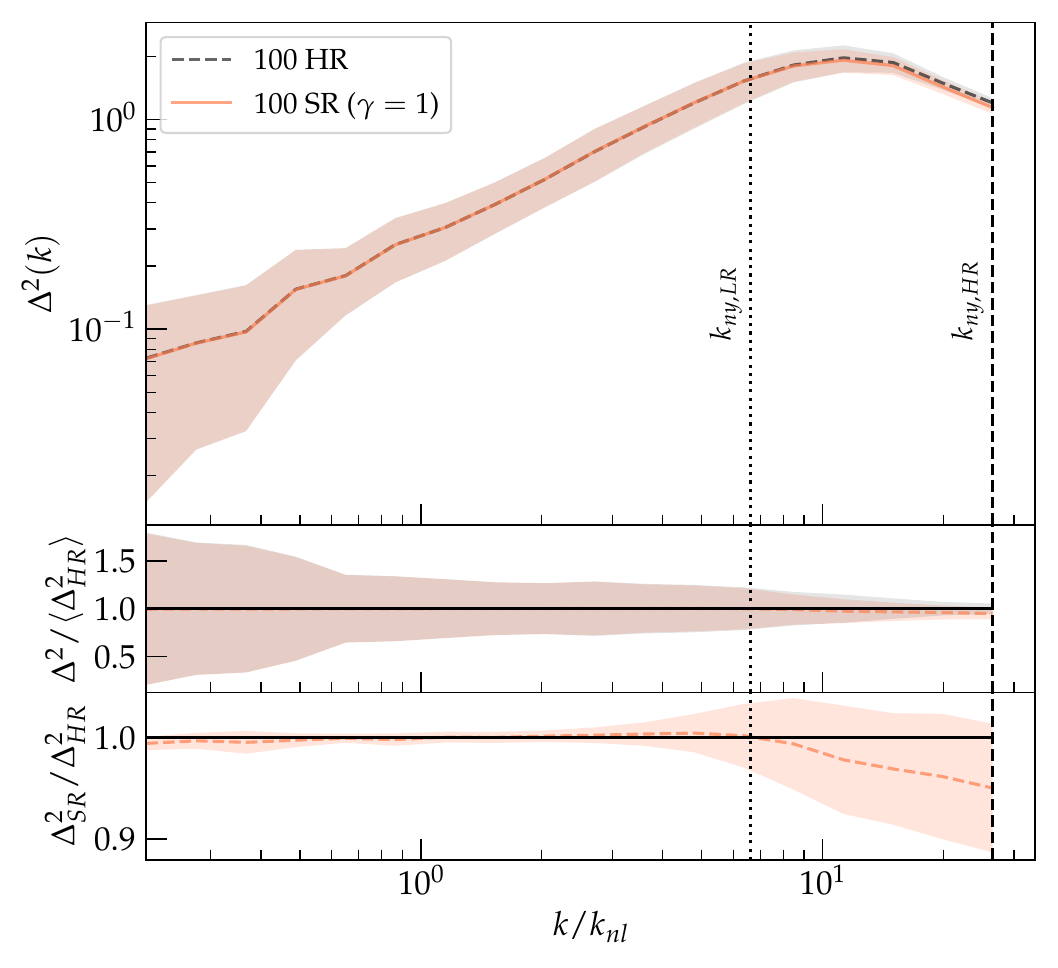}
  \caption[]{\textit{Top panel:} Dimensionless power spectrum of 100 HR and SR simulations based on an ensemble of 100 \textit{different} LR simulations. The mean power is plotted for the HR ensemble (\textit{gray dashed}) and the SR ensemble (\textit{orange}). The 1$\sigma$ deviation around the mean is indicated by the shaded region in the respective color. The HR and SR uncertainty regions overlap exactly at all scales but the smallest ones, with a deviation of less than 5\% for $k > k_{ny, LR}$. \textit{Middle panel:} Ratio between each HR/SR realization and the HR ensemble mean. \textit{Bottom panel:} Ratio between SR and HR; i.e., the reference value is the corresponding HR simulation for each SR realization, rather than the HR ensemble mean.}
  \label{im:100_different}
\end{figure}

For $k = 0.8 \ k_{ny, LR}$, the SR distribution agrees excellently with its HR counterpart. For $k = k_{ny, LR}$ and particularly for $k = k_{ny, HR}$, the width of the SR distribution is also consistent with HR; however, the power is slightly underestimated by our SR model. Generally, the distributions produced by our diffusion model are somewhat more Gaussian than the true HR distributions.

In order to analyze if our diffusion model is able to accurately reproduce the conditional HR distribution not only at the level of the power spectrum, but also pixel-wise, we perform the following coverage test: we consider the (Eulerian) LR density contrast $\delta_{LR}$, flatten it to obtain a list of $64^2$ values between $\delta_{\mathrm{min}} \approx -1$ and $\delta_{\mathrm{max}}$, and determine the pixel indices that correspond to the $0\%$, $12.5\%$, $25\%$, $\ldots$, $100\%$ quantiles. Thus, these pixels are a representative set for the range of density contrasts present in the LR field. We then consider 100 different HR/SR simulations that belong to this LR simulation and, for each HR/SR simulation, extract a single density contrast value from one of the inner $4 \times 4$ subpixels within each of the representative LR pixels. We only consider one of the subpixels (specifically, the HR/SR pixel in the third row and third column within each $4 \times 4$ block) in order to not contaminate ensemble variability, which is the target of this test, with spatial variability.

The resulting HR and SR distributions computed over the 100 realizations are shown in a violin plot in Figure~\ref{im:violin}, for each of the representative pixels. Since these pixels have been selected based on the quantiles of the LR density contrast, one would expect the means of the HR and SR distributions to generally increase when going to pixels that correspond to higher quantiles; however, the HR and SR values will of course depend on the environment.
The distributions for the densest LR pixel (i.e.\ the $100\%$ quantile, given by $\delta_{LR} = 18$) are very wide and would require a significant larger $x$-axis range, for which reason we did not include them in the plot.
The HR and SR distributions agree well across the entire range of densities, both in terms of location and shape, confirming that our diffusion model faithfully reproduces the diversity of HR simulations at the field level. For instance, the uppermost distribution corresponds to a LR pixel with $\delta_{LR} = -1$ (i.e.\ the $0\%$ quantile), and the HR and SR values extracted from a subpixel in 100 realizations both range from $\delta = -1$ to $\sim$0. For LR pixels in denser regions, the HR distributions become much wider: for the last row, $\delta_{LR} = 0.65$ ($87.5\%$ quantile), but negative values of $\delta_{HR}$ occur occasionally in HR subpixels, as well as values $\gtrsim 2$, which is well reproduced by our SR model.

\begin{figure*}[!t]
  \centering
  \includegraphics[width=1\linewidth]{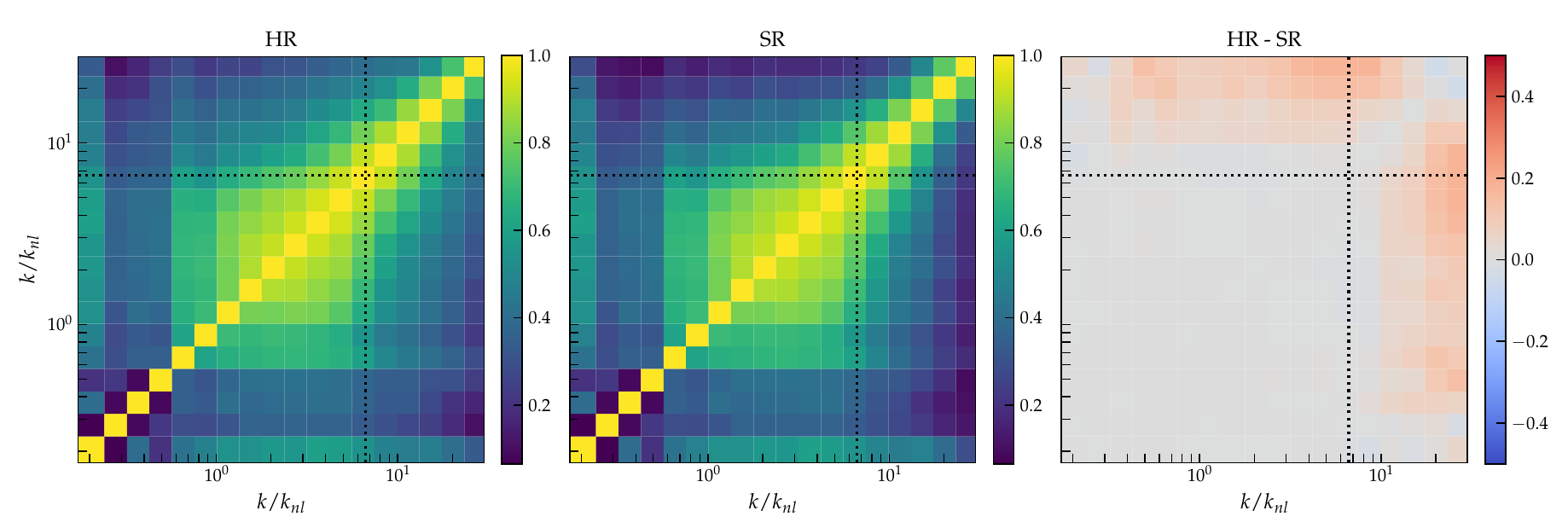}
  \caption[]{Covariance matrix of $\Delta^2(k)$ computed from 1,000 HR (\textit{left}) and SR (\textit{center}) simulations based on 1,000 different LR simulations. The residuals between HR and SR are also presented (\textit{right}). The covariance between scales indicates whether non-linear mode coupling is captured correctly by the SR model. The \textit{dashed} lines indicate the LR Nyquist wave number.}
  \label{im:cov}
\end{figure*}

\subsection{Statistics for many LR images \texorpdfstring{$\mapsto$}{to} many SR images}

Next, we study the SR statistics when marginalizing over different LR realizations. We plot the distribution of the dimensionless power spectrum computed from 100 different HR and SR simulations that belong
to 100 individual LR simulations in Figure~\ref{im:100_different}. 
The shaded regions indicate the 1$\sigma$ scatter over the 
HR (\textit{gray area}) and SR (\textit{orange area}) simulations. 
The distribution of the ratio between each HR/SR realization and the HR ensemble mean is shown in the middle panel. The cosmic variance of the different LR simulations most strongly affects large scales and therefore causes a significant scatter for $k \lesssim k_{ny, LR}$, both for HR and SR. On small scales $k > k_{ny, LR}$, we again observe the trend that our model very slightly underestimates the power.

The bottom panel shows a similar quantity; however, the reference value is now taken to be the corresponding HR simulation \textit{individually for each SR realization}, rather than the HR ensemble mean. Since each HR/SR pair shares the same LR simulation, this eliminates the scatter on large scales due to cosmic variance. While the expected mean of this ratio for a perfect model is one, the slight loss of power on small scales is again visible for our SR samples. Since the single HR simulation taken to be the reference for this panel is not enough to characterize the true distribution of possible HR simulations that belong to the underlying LR simulation, the width of the colored region is less meaningful in this case. The agreement of the power spectra on small scales could potentially be further improved by fine-tuning the model / hyperparameters or the value of $\gamma$ used for the filter.

Finally, in Figure~\ref{im:cov}, we show (normalized) covariance matrices \mbox{$C_{ij} \propto \langle \Delta^2(k_i) \Delta^2(k_j) \rangle$} for the HR simulation, the SR model, and for their difference (from left to right). These covariance matrices capture the degree mode-coupling due to non-linear evolution; for purely linear evolution, the covariance is exactly diagonal. Figure~\ref{im:cov} demonstrates that on the largest scales ($k<k_{nl}$) we still see a predominantly diagonal covariance, as expected on scales that are still linear. In contrast, small scales $k>k_{nl}$ are strongly coupled. The SR model captures both of these features qualitatively very well. 

The residual (right panel) indicates however that non-linear mode-coupling in the regime $k>10 k_{nl}$ (increasing close to the HR Nyquist mode) is however somewhat underestimated in the SR model, while the terms closer to the diagonal are captured accurately. This finding is consistent with the slight bias in the dimensionless power spectrum at the HR Nyquist mode that was already found in Figure~\ref{im:sigma_distribution}. Our current SR model therefore consistently underestimates the non-linear mode coupling close to the HR Nyquist mode at the level of 10$-$20 percent, which could potentially be improved by giving even more weight to small scales in the filter boosted training, which is however beyond the scope of this study.

\section{Conclusions}\label{sec:conclusions}

We demonstrated that our diffusion model, together with filter-boosted training, can generate convincing SR simulations which are, on a statistical level, in high agreement with the original HR realizations. Filter-boosted training provides a straightforward means to improve accuracy at scales that are otherwise difficult to reproduce by a pixel-wise loss. Not only does the power spectrum match up to $>$5\% at scales $k > k_{ny, LR}$ (same range as in \citealt{Li_2021} and \citealt{Sipp2022}), our model is able to perform \textit{stochastic} super-resolution simulations and provide meaningful uncertainties. Our results indicate that the covariance between scales is a sensitive indicator for the performance of such models.

In the context of cosmological simulations, our diffusion-based super-resolution framework can be further expanded on. For instance, \cite{Zhang2023} introduced a redshift dependence to their super-resolution GAN model. An additional extension would be to make our model cosmology dependent, meaning that super-resolution could be performed for different values of e.g.\ $\Omega_m$ and $\sigma_8$, together with meaningful uncertainties. The scale-free nature of our set-up should in principle allow arbitrarily high super-resolution when training on simulations covering a large range of different times and length scales -- an aspect we plan to study in future work.

An interesting extension would be to conduct a full investigation for 3D simulations. This faces the difficulty that diffusion models are expensive to train and even during inference need multiple minutes on a current state-of-the-art GPU to sample with multiple diffusion steps. To improve this, one could for example utilize a denoising implicit model \citep{song2022denoising} or other recent improvements that cut steps of the Markov chain during inference (see e.g. \citealt{nichol2021improved}).

Filter-boosted training is also applicable to other generative tasks beyond a cosmological setting. It would be interesting to study classical machine learning tasks such as the reconstruction of faces and to analyze whether filter-boosted training has a significant impact on standard metric scores (e.g. FID).

A challenge of filter-boosted training is the choice of the filter and the strength (i.e. value of $\gamma$). Especially, the search for a suitable $\gamma$-value can become computationally expensive. Therefore, \textit{trainable} filters are a promising avenue, where the `strength' of the filter becomes a learnable parameter during training, represented by an additional term in the loss function. 

{\vspace{0.5em} \small \it While this paper was undergoing peer review, a preprint was posted on arXiv by \citet{rouhiainen2023superresolution}, who showcase in a similar technical setup how an iterative procedure can be used for super-resolving 3D cosmological simulation boxes with diffusion models.}

\section*{Acknowledgements}

The computational results presented have been achieved using the Vienna Scientific Cluster (VSC), specifically the VSC5.

\section*{Data Availability}
\noindent The trained super-resolution model is available from the authors upon reasonable request.



\bibliographystyle{mnras}
\bibliography{paper.bib} 

\setlength{\twocolwidth}{\columnwidth}




\end{document}